\documentclass[%
 reprint,
showpacs,preprintnumbers,
 amsmath,amssymb,
 aps,
pra,
]{revtex4-1}

\usepackage[colorinlistoftodos]{todonotes}
\usepackage{graphicx}% Include figure files
\usepackage{dcolumn}% Align table columns on decimal point
\usepackage{bm}% bold math
\begin{document}

%\preprint{APS/123-QED}

\title{Perturbation theory for graphene integrated waveguides: \\cubic nonlinearity and third harmonic generation }% Force line breaks with \\

\author{Andrey V. Gorbach}
\email{A.Gorbach@bath.ac.uk}
\author{Edouard Ivanov}%
 \affiliation{%
 Centre for Photonics and Photonic Materials \\
 Department of Physics, University of Bath, Bath BA27AY, UK
}%

\date{\today}% It is always \today, today,

\begin{abstract}
We present perturbation theory for analysis of generic third-order nonlinear processes in graphene integrated photonic structures. Optical response of graphene is treated as the nonlinear boundary condition in Maxwell equations. The derived models are applied for analysis of third harmonic generation in a graphene coated dielectric micro-fibre. The efficiency of up to few percent is predicted when using sub-picosecond pump pulses with energies of the order of $0.1$nJ in a sub-millimeter long fibre, when operating near the resonance of the graphene nonlinear conductivity $\hbar\omega=(2/3)E_F$.
\end{abstract}

\maketitle

%\tableofcontents

\section{\label{sec:intro}Introduction}

Optical and opto-electonic properties of graphene, accustomed to its unique electron dispersion in a vicinity of the so-called Dirac cones, have been in focus of intensive research recently \cite{Bonaccorso2010, Bao2012}. In particular, this linear (i.e. massless) band structure has been identified as the origin of the exceptionally strong nonlinear optical response of graphene \cite{Mikhailov2007, Mikhailov2008, Glazov2014, Mikhailov2015, Cheng2015}. Several experimental studies confirmed that the effective third-order nonlinear coefficient $\chi_3$ (Kerr coefficient) of graphene exceeds that of typical dielectrics by six to eight orders of magnitude \cite{Hendry2010,Zhang2012,Hong2013,Kumar2013}. These discoveries make graphene the particularly attractive material for integration with various nonlinear photonic components, such as waveguides and cavities. Indeed, a considerable boost of third-order nonlinear processes has been demonstrated in a graphene-coated photonic crystal cavity \cite{Gu2012}, a photonic crystal waveguide \cite{Zhou2014}, and silica micro-fibres \cite{Gao2014, Wu2015}.

Being purely two-dimensional structure, graphene is conceptually different from any bulk material. When using conventional theoretical tools to describe nonlinear processes in a graphene integrated structure, one is compelled to treat graphene as a thin film with certain {\em bulk} linear and nonlinear dielectric constants \cite{Gu2012, Wu2015, Donnelly2014, Nesterov2012}. This can only be justified for setups where electric field is polarized in the plane of graphene. However, in micro- and nano-metre size photonic and plasmonic structures, where the typical localization scale of guided/cavity modes is comparable to (or smaller than) the wavelength, all vector components of the electric field can be strongly pronounced.

An alternative approach is to treat graphene as the surface current boundary condition. 
This method appears to be more adequate when dealing with one atom thick materials, and proves to give accurate description of graphene surface plasmons \cite{Jablan2009, Koppens, Xiao2016}.
Considering nonlinear optical response of graphene, this approach implies introduction of the corresponding nonlinear boundary condition in Maxwell equations, in addition to nonlinear polarization terms describing bulk materials. Recently we developed the corresponding perturbation expansion procedure of Maxwell equations to describe self-focusing and switching of monochromatic graphene surface plasmons in single- and bi-layer graphene structures \cite{Gorbach2013,Smirnova2013}, as well as self-phase modulation and nonlinear frequency broadening of pulses propagating in graphene-coated dielectric fibres \cite{Gorbach2013ol} and graphene plasmonic waveguides \cite{Gorbach2015}.

\begin{figure}
\begin{center}
\includegraphics[width=0.4\textwidth]{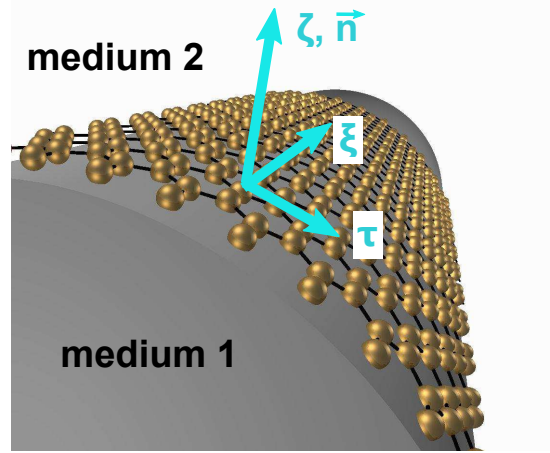}
\end{center}
\caption{(Color online) Graphene induced surface current: local coordinates.}
\label{fig:bound_cond}
\end{figure}

In this work we extend the procedure onto the generic problem of nonlinear frequency mixing, including third harmonic generation, in graphene integrated optical waveguides. It is assumed that at each frequency component of the signal, the structure supports a discrete set of linear guided modes. The guidance is provided either exclusively by bulk dielectrics (as in graphene-coated dielectric fibres), or by the integrated graphene sheet (as in graphene plasmonic waveguides). In the former case, the total graphene induced surface current is treated as a perturbation, together with the nonlinear polarization terms originating from bulk materials. In the latter case, the linear part of the surface current is incorporated in the guided mode analysis, while the nonlinear current and polarization terms are treated as perturbations. 

The general procedure of perturbation expansion of Maxwell equations with nonlinear polarization and nonlinear surface current terms is described in Section \ref{sec:ii}. In Section \ref{sec:iii} this procedure is applied for the case of third-order nonlinearities, and the corresponding set of coupled differential equations describing evolution of modal amplitudes with the propagation distance is derived in frequency domain. In Section \ref{sec:thg} we consider the specific problem of third harmonic generation from a relatively narrow band-width pump, assuming that the phase matching condition is satisfied for one particular pair of pump and third harmonic modes of the waveguide. For this case, the modal equations are reduced to the conventional system of two coupled nonlinear Schr{\"o}dinger type equations in time domain. To illustrate the application of derived models, in Section \ref{sec:thg_fibre} we analyze third harmonic generation in a graphene-coated silica micro-fibre. In particular, we consider the conversion efficiency and the optimal length of the graphene coated section. Also, we compare graphene induced changes to modal propagation constants (including attenuation constants) in the fundamental and third harmonics as predicted by the perturbation theory and computed directly with the help of the commercial finite element method Maxwell solver package Comsol Multiphysics.

%%%%%%%%%%%%%%%%%%%%%%%%%%%%%%%%%%%%%%%%%%%%%%%%%%%%%%%%%%%%%
\section{Perturbation expansion of Maxwell equations} 
\label{sec:ii}

Consider a graphene integrated waveguide with a fixed cross-section along the propagation direction $z$. To  describe nonlinear wave propagation in the structure, it is convenient to use Fourier expansion of the real electric field:
\begin{equation}
\vec{\mathcal{E}}(\vec{r},t)=\frac{1}{2\sqrt{2\pi}}\int_{0}^{+\infty}\mathbf{E}(\vec{r},\omega)e^{-i\omega t}d\omega + c.c.\;,
\label{eq:el_field_gen_ansatz}
\end{equation}
and similar expansions of other fields. Here, $\mathbf{E}$ is assumed to be a vector function of {\em positive only} frequencies.

Each Fourier component $\mathbf{E}$ solves Maxwell equations:
\begin{equation}
\vec{\nabla}\times\vec{\nabla}\times \mathbf{E}= \frac{\omega^2}{c^2\epsilon_0}\mathbf{D}\;.
\label{eq:Maxwell}
\end{equation}
Optical response of all bulk materials is incorporated in the displacement vector $\mathbf{D}$. Atom-thick graphene layer is described by means of the surface current $\mathbf{J}$, the corresponding boundary condition is:
\begin{equation}
\vec{n}\times\left[\mathbf{H}_2-\mathbf{H}_1\right]=\mathbf{J}\;,
\end{equation}
where $\vec{n}$ is the unit vector normal to the graphene layer and pointing from medium $1$ to medium $2$, which are on either side of the graphene layer, see Fig.~\ref{fig:bound_cond}. Introducing local coordinates $(\xi,\tau,\zeta)$, where $\zeta$ is orthogonal to the graphene layer (the layer is located at $\zeta=0$) and $(\xi,\tau)$ are in-plane of graphene, as shown in Fig.~\ref{fig:bound_cond}, the above boundary condition can be written as:
\begin{eqnarray}
\label{eq:BCs}
\Delta\left[H_\xi\right]=J_\tau\;, \qquad
\Delta\left[H_\tau\right]=-J_\xi\;.
\end{eqnarray} 
Here, operator $\Delta$ is defined as:
\begin{eqnarray}
\label{eq:op_Delta}
\Delta[f(\zeta)]&=&\lim_{\delta\to 0}\left(f(-\delta)-f(\delta)\right)\;,
\end{eqnarray} 
and characterizes the variation of a function $f(\zeta)$ across the graphene boundary.

It is convenient to decompose the displacement vector and the induced current as $\mathbf{D}=\mathbf{D}_l+\mathbf{D}_p$, $\mathbf{J}=\mathbf{J}_l+\mathbf{J}_p$, so that solution of Maxwell equations with $\mathbf{D}_l$ and $\mathbf{J}_l$ gives {\em linear} guided modes of the structure, while $\mathbf{D}_p$ and $\mathbf{J}_p$ are treated as perturbations and contain third-order nonlinear terms. Thus for $\mathbf{D}_l$ and $\mathbf{J}_l$ we assume:
\begin{eqnarray}
\label{eq:Dl}
\mathbf{D}_l&=&\epsilon_0\epsilon\mathbf{E}\;,\\
\mathbf{J}_{l}&=&\hat{\sigma}_l\mathbf{E}\;,
\label{eq:Jl}
\end{eqnarray}
where $\epsilon=\epsilon(\mathbf{r}_\perp)$ is the relative dielectric permittivity, $\mathbf{r}_\perp$ is the subset of coordinates orthogonal to the propagation direction $z$, $\hat{\sigma}_l$ is linear conductivity tenzor which ensures that current $\mathbf{J}_l$ has only in-plane components, i.e. $J_{l,\zeta}=0$ cf. Fig.~\ref{fig:bound_cond}. 

Below we assume that $\hat{\sigma}_l$ is purely imaginary: $\hat{\sigma}_l^*=-\hat{\sigma}_l$, so that linear guided modes are lossless.
For some graphene-integrated photonic structures, such as e.g. graphene-coated photonic crystal cavities \cite{Gu2012} and micro-fibres \cite{Gorbach2013ol}, the linear guided mode is supported exclusively by the bulk structure, while the additional graphene layer introduces only minor corrections, cf. Fig.~\ref{fig:fibre_charact}.
In such case it is reasonable to set $\hat{\sigma}_l=0$, and keep total graphene surface current (linear and non-linear) in the perturbation term $\mathbf{J}_p$. On the contrary, for graphene plasmonic waveguides \cite{Gorbach2015} linear graphene conductivity defines the structure of guided modes. Here, it is essential to keep imaginary part of linear graphene conductivity in the leading perturbation expansion order, while real part (which gives damping of plasmons) is included in $\mathbf{J}_p$ (typically, $\textrm{Re}(\sigma)/\textrm{Im}(\sigma)\ll 1$ for graphene plasmons \cite{Yan2013b}).  
%While it is possible to include linear damping (real part of linear graphene conductivity) in the leading order  $\mathbf{J}_l$ and consider quasi-guided modes with complex propagation constants, we find it more convenient to work with real guided modes.

Developing perturbation expansion, we introduce a dummy small parameter $s$, assuming $\mathbf{D}_p,\mathbf{J}_p\sim s^{3}$. Each Fourier component of the electric field is expanded in the perturbation series as:
\begin{eqnarray}
\nonumber
\mathbf{E}&=&\sum_j\left\{s \frac{A_{\omega,j}(s^2z)}{\sqrt{N_{\omega,j}}}\mathbf{e}_{\omega,j}(\mathbf{r}_\perp)+s^{3}\mathbf{B}_{\omega,j}(\mathbf{r}_\perp,s^2 z)\right\}e^{i\beta_j z}\\
&&+O(s^{5})\;,
\label{eq:expansion_ansatz}
\end{eqnarray}
and a similar expansion for the magnetic field is assumed. Here $\mathbf{e}_{\omega,j}$ is a $j$-th linear mode of the structure, $\beta_j=\beta_j(\omega)$ is the corresponding propagation constant,  $N_{\omega,j}$ is an optional normalization factor. 

In other words, the cumulative effect of perturbations in polarization $\mathbf{D}_p$ and surface current $\mathbf{J}_p$ is sought in the form of slow variation (on the scale of the wave period $2\pi/\beta_j$) of modal amplitudes $A_j$ with the propagation distance $z$, and corrections to the shape of the modes $\mathbf{B}$. The particular hierarchy of powers of the small parameter $s$ in Eq. (\ref{eq:expansion_ansatz}) is specific for third-order nonlinearities \cite{Gorbach2013,Marini2011}, it is justified below by consistently solving boundary value problems, which emerge in different orders of $s$.

Following substitution of the ansatz in Eq.~(\ref{eq:expansion_ansatz}) into Maxwell equations, in the lowest order of the small parameter, $O(s)$, the eigenvalue problem is obtained:
\begin{eqnarray}
\label{eq:ev_Os12}
\hat{L}(\beta)\mathbf{e}_{\omega}=0\;, 
\end{eqnarray}
where operator $\hat{L}(\beta)$ is defined as:
\begin{eqnarray}
\hat{L}\mathbf{e}=e^{-i\beta z}\left\{\vec{\nabla}\times\vec{\nabla}\times\mathbf{e}(\mathbf{r}_\perp)e^{i\beta z}\right\}
-\frac{\omega^2}{c^2}\epsilon\mathbf{e}(\mathbf{r}_\perp)\;,
\label{eq:operator_L}
\end{eqnarray}
Solving this eigenvalue problem, we obtain a set of modal profiles $\mathbf{e}_{\omega,j}(\mathbf{r}_\perp)$ together with the propagation constants $\beta_j(\omega)$.

We choose the normalization factors $N_{\omega,j}$ via the orthogonality condition of guided modes:
\begin{eqnarray}
\frac{1}{4}\int (\mathbf{e}_{\omega,j} \times \mathbf{h}_{\omega,k}^*+\mathbf{e}_{\omega,k}^* \times \mathbf{h}_{\omega,j})\hat{e}_z d\Omega=N_{\omega,j}\delta_{jk}\;,
\label{eq:Iw}
\end{eqnarray}
where $d\Omega$ is the unit area and integration is performed over the entire cross-section of the waveguide, $\hat{e}_z$ is the unit vector along $z$-axis, $\delta_{jk}$ is the Kronecker's delta. It is easy to see that with such normalization, in the lowest order of the small parameter $O(s^2)$, the total energy carried by a pulse along the waveguide is given by:
\begin{equation}
W=\iint_{-\infty}^{\infty}\left(\mathcal{E}\times\mathcal{H}\right)\hat{e}_z d\Omega dt=\sum_j\int_0^{\infty}|A_{\omega,j}|^2d\omega\;.
\label{eq:energy_W}
\end{equation}

In the next order of the perturbation expansion of Maxwell equations, $O(s^{3})$, we obtain:
\begin{eqnarray}
\nonumber
\sum_j\left\{\hat{L}\mathbf{B}_{\omega,j}- \frac{1}{\sqrt{N_{\omega,j}}}\partial_z A_{\omega,j}
\hat{M}\mathbf{e}_{\omega,j}\right\}e^{i\beta_jz}=\\
\frac{\omega^2}{c^2\epsilon_0}\mathbf{D}_p\;.\;
\label{eq:Os32}
\end{eqnarray}
The structure of operator $\hat{M}$ is specified in the Appendix using Cartesian and cylindrical coordinates, see Eqs. (\ref{eq:M_cart}) and (\ref{eq:M_cyl}), respectively.

Next, we project Eq.~(\ref{eq:Os32}) onto the mode $\mathbf{e}_{\omega,k}$ using the following scalar product definition:
\begin{equation}
\left<\mathbf{a} |\mathbf{b}\right>=\int (\mathbf{a}^*\cdot \mathbf{b})d\Omega\;.
\end{equation}

It is important to note that $\mathbf{e}_\omega$ and $\mathbf{B}_\omega$ satisfy different boundary conditions in Eq.~(\ref{eq:BCs}) by virtue of the earlier introduced separation of the total current into the leading order $\mathbf{J}_l$ and perturbation $\mathbf{J}_p$ parts. This removes the self-adjoint property of the operator $\hat{L}$, so that $\left<\mathbf{e}_\omega|\hat{L}|\mathbf{B}_\omega\right>\ne\left<\mathbf{B}_\omega|\hat{L}|\mathbf{e}_\omega\right>^*$, cf. Eq.~(\ref{eq:L_not_adjoint}) in the Appendix. Constructing projections on both sides of Eq.~(\ref{eq:Os32}) and taking some components of the resulting integrals by parts, adopting the definition of eigenmodes in Eq.~(\ref{eq:ev_Os12}), applying boundary conditions in Eq.~(\ref{eq:BCs}), and using the linearity of $\mathbf{J}_l(\mathbf{E})$ in Eq.~(\ref{eq:Jl}), it is possible to derive (see Appendix for details):
\begin{eqnarray}
\nonumber
&&\sum_je^{i\beta_j z}\frac{-i\partial_z A_{\omega,j}}{\sqrt{N_{\omega,j}}}\int
(\mathbf{e}_{\omega,j} \times \mathbf{h}_{\omega,k}^*+\mathbf{e}_{\omega,k}^* \times \mathbf{h}_{\omega,j})\hat{e}_zd\Omega\\
&&\qquad=i\int_C\left(\mathbf{e}_{\omega,k}^*\cdot\mathbf{J}_p\right)dl
+\omega\int\left(\mathbf{e}_{\omega,k}^*\cdot\mathbf{D}_p\right)d\Omega\;,
\label{eq:after_appendix}
\end{eqnarray}
where $\int_C\left<\dots\right>dl$ is line integral along the contour of the graphene sheet introduced in the cross-sectional plane of the waveguiding structure, cf. Fig.~\ref{fig:bound_cond}. 

Finally, using the normalization condition in Eq.~(\ref{eq:Iw}), we obtain the following equation:
\begin{eqnarray}
\nonumber
&&\partial_z A_{\omega,k}=\frac{e^{-i\beta_kz}}{4\sqrt{N_{\omega,k}}}\times\\
&&\qquad\left[
i\omega\int\left(\mathbf{e}_{\omega,k}^*\cdot\mathbf{D}_p\right)d\Omega
-\int_C\left(\mathbf{e}_{\omega,k}^*\cdot\mathbf{J}_p\right)dl
\right]\;,
\label{eq:modal}
\end{eqnarray}
which describes the evolution of modal amplitudes with propagation distance, induced by perturbations in bulk polarization and surface (graphene) current.

\section{Nonlinear polarization and current}
\label{sec:iii}

In this work we focus on cubic nonlinearity of bulk media and graphene. The corresponding perturbation polarization and current can be written as:
\begin{eqnarray}
\label{eq:Dpreal}
\vec{\mathcal{D}}_p=\epsilon_0\hat{\chi}^{(3)}\vdots \vec{\mathcal{E}}^3\;,\\
\label{eq:Jpreal}
\vec{\mathcal{J}}_p=\hat{\sigma}_R\vec{\mathcal{E}}+\hat{\sigma}^{(3)}\vdots \vec{\mathcal{E}}^3\;,
\end{eqnarray}
where vertical dots stand for tensor product: $\mathbf{a}=\hat{O}\vdots \mathbf{b}\mathbf{c}\mathbf{d}$, $a_i=\hat{O}_{ijkl}b_j c_k d_l$. We keep linear term in the perturbation current to account for graphene-induced losses and (in case of  $\hat{\sigma}_l=0$) small corrections to propagation constants. Assuming no anisotropy in the graphene plane (e.g. due to external magnetic fields), linear tensor $\hat{\sigma}_R$ has only two non-zero components on the diagonal: $\hat{\sigma}_{R,\xi\xi}=\hat{\sigma}_{R,\tau\tau}=\sigma_R+i\sigma_I$, where $\sigma_R$ and $\sigma_I$ are real constants, $\sigma_R>0$.

Substituting the Fourier expansion for electric field, Eq.~(\ref{eq:el_field_gen_ansatz}), into Eqs.~(\ref{eq:Dpreal}), (\ref{eq:Jpreal}) , and
adopting similar expansions for $\mathcal{D}_p$ and $\mathcal{J}_p$, we thus obtain the corresponding expressions for Fourier components of the perturbation fields:

%\begin{widetext}

\begin{eqnarray}
\nonumber
&&\mathbf{D}_p=\frac{\epsilon_0}{8\pi}\iint d\omega_1d\omega_2\left\{\hat{\chi}^{(3)}\vdots\mathbf{E}(\omega_1)\mathbf{E}(\omega_2)\mathbf{E}(\omega_{+-})+\right.\\
\nonumber
&&\qquad\qquad 
3\hat{\chi}^{(3)}\vdots\mathbf{E}(\omega_1)\mathbf{E}^*(\omega_2)\mathbf{E}(\omega_{++})+\\
\label{eq:Dp}
&&\qquad \qquad \left.
3\hat{\chi}^{(3)}\vdots\mathbf{E}(\omega_1)\mathbf{E}^*(\omega_2)\mathbf{E}^*(\omega_{-+})\right\}\;,\\
\nonumber
&&\mathbf{J}_p=\hat{\sigma}_R\mathbf{E}(\omega)+\\
\nonumber
&&\qquad\frac{1}{8\pi}\iint d\omega_1d\omega_2\left\{\hat{\sigma}^{(3)}\vdots\mathbf{E}(\omega_1)\mathbf{E}(\omega_2)\mathbf{E}(\omega_{+-})+\right.\\
\nonumber
&&\qquad\qquad 
3\hat{\sigma}^{(3)}\vdots\mathbf{E}(\omega_1)\mathbf{E}^*(\omega_2)\mathbf{E}(\omega_{++})+\\
\label{eq:Jp}
&&\qquad\qquad 
\left.3\hat{\sigma}^{(3)}\vdots\mathbf{E}(\omega_1)\mathbf{E}^*(\omega_2)\mathbf{E}^*(\omega_{-+})\right\}\;,\\
\label{eq:omegapm}
&&\omega_{+-}=\omega-\omega_1-\omega_2\;,\\
\label{eq:omegapp}
&&\omega_{++}=\omega-\omega_1+\omega_2\;,\\
\label{eq:omegamp}
&&\omega_{-+}=-\omega-\omega_1+\omega_2\;.
\end{eqnarray}
%\end{widetext}
Generally, the nonlinear tensors in the above integrals are functions of the three frequencies of the vectors their acting upon: $\hat{\chi}^{(3)}=\hat{\chi}^{(3)}(\omega_1,\omega_2,\omega_{\pm\pm})$, $\hat{\sigma}^{(3)}=\hat{\sigma}^{(3)}(\omega_1,\omega_2,\omega_{\pm\pm})$. 

The requirement for all frequencies to be positive introduces certain selection rules in the above integrals, resulting in different integration limits set for different integrant parts, cf. Eqs.~(\ref{eq:omegapm})-(\ref{eq:omegamp}). Instead, it is often convenient to introduce the extension of $\mathbf{E}(\omega)$ and all other field functions onto domain of negative frequencies, assuming that all fields are zero in that domain: $\mathbf{E}(\omega<0)\equiv 0$. In other words, we assume that the Fourier transform of real electric field can be decomposed as: $\mathcal{F}(\vec{\mathcal{E}}(t))_\omega=\mathbf{E}(\omega)+\mathbf{E}^*(-\omega)$ with $\mathbf{E}(\omega)$ now defined on the entire real axis of frequencies. This is equivalent to separation into forward- and backward-propagating modes, and it is the common approach for analysis of nonlinear guided waves \cite{agrawal,Kolesik2004a}. Adopting this approach, hereafter we can set the integration limits in r.h.s. of Eqs.~(\ref{eq:Dp}) and ~(\ref{eq:Jp}) to cover the entire real frequencies domain.

For isotropic homogeneous media, the general structure of the third-order susceptibility tensor is given by \cite{boyd}:
 \begin{equation}
 \hat{\chi}^{(3)}_{ipjs}=\frac{\chi_3}{3}
 \left[\delta_{ip}\delta_{js}+\delta_{ij}\delta_{ps}+\delta_{is}\delta_{pj}\right]\;,
 \end{equation}
 where $\chi_3=\hat{\chi}^{(3)}_{xxxx}$. The third-order conductivity tensor has the same structure for the subset of coordinates in the graphene plane $(\xi,\tau)$ \cite{Mikhailov2015}. Also, the 2D symmetry of graphene and the assumption of zero transverse current $J_\zeta=0$ 
still permit six additional non-zero tensor components $\hat{\sigma}^{(3)}_{jj\zeta\zeta}=\hat{\sigma}^{(3)}_{j\zeta j\zeta}=\hat{\sigma}^{(3)}_{j\zeta\zeta j}=\widetilde{\sigma_3}$, $j=\xi,\tau$. Thus the nonlinear conductivity tensor can be written as follows:
\begin{eqnarray}
\nonumber
\hat{\sigma}^{(3)}_{ipjs}&=&\frac{\sigma_3}{3}
\left[\delta_{ip}\delta_{js}+\delta_{ij}\delta_{ps}+\delta_{is}\delta_{pj}\right] \left[\delta_{i\xi}+\delta_{i\tau}\right]\\
&&\times\left[1+(\nu-1)(\delta_{j\zeta}\delta_{s\zeta}+\delta_{p\zeta}\delta_{s\zeta}+\delta_{p\zeta}\delta_{j\zeta})\right]
\;,\qquad
\end{eqnarray}
where $\sigma_3=\hat{\sigma}^{(3)}_{xxxx}$ and $\nu=3\widetilde{\sigma_3}/\sigma_3$.

With the above structure of third-order tensors, it is easy to see that:
\begin{eqnarray}
\label{eq:chi3_via_scalar}
\hat{\chi}^{(3)}\vdots \mathbf{a}\mathbf{b}\mathbf{c}&=&\frac{\chi_3}{3}\left[
\left(\mathbf{a}\cdot\mathbf{b}\right)\mathbf{c}+
\left(\mathbf{b}\cdot\mathbf{c}\right)\mathbf{a}+
\left(\mathbf{a}\cdot\mathbf{c}\right)\mathbf{b}
\right]\;,\qquad\\
\nonumber
\left(\hat{\sigma}^{(3)}\vdots \mathbf{a}\mathbf{b}\mathbf{c}\right)_i&=&\delta_{i\xi}\delta_{i\tau}\frac{\sigma_3}{3}\times\\
\label{eq:sig3_via_scalar}
&&\left[
\left(\mathbf{a}\cdot\mathbf{b}\right)_\nu\mathbf{c}+
\left(\mathbf{b}\cdot\mathbf{c}\right)_\nu\mathbf{a}+
\left(\mathbf{a}\cdot\mathbf{c}\right)_\nu\mathbf{b}
\right]\;,\qquad
\end{eqnarray}
with the deformed scalar product defined as: 
\begin{equation}
(\mathbf{a}\cdot\mathbf{b})_\nu=\nu a_\zeta b_\zeta+a_\xi b_\xi+a_\tau b_\tau\;.
\label{eq:deform_scalar_product}
\end{equation}
Note, in the limit of a monochromatic wave $\mathbf{E}(\omega)=\sqrt{2\pi}\delta(\omega-\omega_0)\mathbf{E}_{cw}$, $\mathbf{D}(\omega)=\sqrt{2\pi}\delta(\omega-\omega_0)\mathbf{D}_{cw}$, Eqs.~(\ref{eq:Dl}), (\ref{eq:Dp}) and (\ref{eq:chi3_via_scalar}) give the conventional relationship $\mathbf{D}_{cw}=\epsilon_0\epsilon \mathbf{E}_{cw}+(\epsilon_0 \chi_3/4)\left(2|\mathbf{E}_{cw}|^2\mathbf{E}_{cw}+\mathbf{E}_{cw}^2\mathbf{E}_{cw}^*\right)$ \cite{boyd}.

Substituting expressions for the perturbation current and polarization from Eqs.~(\ref{eq:Dp}), (\ref{eq:Jp}) into Eq.~(\ref{eq:modal}) , the following set of first order coupled nonlinear differential equations for modal amplitudes $A_{\omega,k}(z)$ is obtained:
\begin{eqnarray}
\nonumber
&&\partial_z A_{\omega,k}=\left(i\alpha_{\omega,k}-\kappa_{\omega,k}\right) A_{\omega,k} +\\
\nonumber
&&\;\;
\sum_{i,p,s}\frac{i}{2\pi}\iint_{-\infty}^{+\infty}\left\{\gamma_{kips}^{(+-)}A_{\omega_1,i}A_{\omega_2,p}
A_{\omega_{+-},s}e^{i\Delta\beta_{+-}z}\right.\\
\nonumber
&&\qquad
+3\gamma_{kips}^{(++)}A_{\omega_1,i}A^*_{\omega_2,p}
A_{\omega_{++},s}e^{i\Delta\beta_{++}z}\\
\label{eq:modal_coupled}
&&\qquad
\left.
+3\gamma_{kips}^{(-+)}A_{\omega_1,i}A^*_{\omega_2,p}
A_{\omega_{-+},s}^*e^{i\Delta\beta_{-+}z}
\right\}d\omega_1d\omega_2\;,\;\;
\end{eqnarray}
where:
\begin{eqnarray}
&&\Delta\beta_{+-}=\beta_i(\omega_1)+\beta_p(\omega_2)+\beta_s(\omega_{+-})-\beta_k(\omega)\;,
\label{eq:Delta_beta_pm}\\
&&\Delta\beta_{++}=\beta_i(\omega_1)-\beta_p(\omega_2)+\beta_s(\omega_{++})-\beta_k(\omega)\;,
\label{eq:Delta_beta_pp}\\
&&\Delta\beta_{-+}=\beta_i(\omega_1)-\beta_p(\omega_2)-\beta_s(\omega_{-+})-\beta_k(\omega)\;,
\label{eq:Delta_beta_mp}
\end{eqnarray}
the graphene induced complex corrections to the propagation constants are given by:
\begin{equation}
\kappa_{\omega,k}-i\alpha_{\omega,K}=\frac{1}{4N_{\omega,k}}\int_C \left(\sigma_R+i\sigma_I\right) (e_{\omega,k}^*\cdot e_{\omega,k})_0dl\;,
\label{eq:kappa}
\end{equation}
the modified scalar product $(\mathbf{a}\cdot\mathbf{b})_0$ takes into account only components of vectors in the graphene plane, i.e. it corresponds to $\nu=0$ in the earlier defined product in Eq.~(\ref{eq:deform_scalar_product}), and nonlinear coefficients combine contributions from bulk dielectric polarization and graphene surface current:
\begin{eqnarray}
&&\gamma_{kips}^{(\mu\nu)}=\frac{1}{16\sqrt{N_1 N_2 N_3 N_4}}
\left[
\epsilon_0\omega\Gamma_{kips}^{(\mu\nu,d)}+i\Gamma_{kips}^{(\mu\nu,g)}
\right]\;,\\
\label{eq:gam_d_pm}
&&\Gamma_{kips}^{(+-,d)}=
\int\chi_3S(\mathbf{e}_2,\mathbf{e}_3,\mathbf{e}_4;\mathbf{e}_1^*)d\Omega\;,\\
\label{eq:gam_d_pp}
&&\Gamma_{kips}^{(++,d)}=
\int\chi_3S(\mathbf{e}_2,\mathbf{e}_3^*,\mathbf{e}_4;\mathbf{e}_1^*)d\Omega\;,\\
\label{eq:gam_d_mp}
&&\Gamma_{kips}^{(-+,d)}=
\int\chi_3S(\mathbf{e}_2,\mathbf{e}_3^*,\mathbf{e}_4^*;\mathbf{e}_1^*)d\Omega\;,\\
\label{eq:gam_gr_pm}
&&\Gamma_{kips}^{(+-,g)}=
\int_C\sigma_3\widetilde{S}(\mathbf{e}_2,\mathbf{e}_3,\mathbf{e}_4;\mathbf{e}_1^*)dl\;,\\
\label{eq:gam_gr_pp}
&&\Gamma_{kips}^{(++,g)}=
\int_C\sigma_3\widetilde{S}(\mathbf{e}_2,\mathbf{e}_3^*,\mathbf{e}_4;\mathbf{e}_1^*)dl\;,\\
\label{eq:gam_gr_mp}
&&\Gamma_{kips}^{(-+,g)}=
\int_C\sigma_3\widetilde{S}(\mathbf{e}_2,\mathbf{e}_3^*,\mathbf{e}_4^*;\mathbf{e}_1^*)dl\;,\\
\nonumber
&&S(\mathbf{a},\mathbf{b},\mathbf{c};\mathbf{d})=\frac13\left[
(\mathbf{a}\mathbf{b})(\mathbf{c}\mathbf{d})+
(\mathbf{a}\mathbf{c})(\mathbf{b}\mathbf{d})
\right.\\
\label{eq:S}
&&\qquad\qquad\qquad\left.+(\mathbf{b}\mathbf{c})(\mathbf{a}\mathbf{d})
\right]\;,\\
\nonumber
&&\widetilde{S}(\mathbf{a},\mathbf{b},\mathbf{c};\mathbf{d})=\frac13\left[
(\mathbf{a}\mathbf{b})_\nu(\mathbf{c}\mathbf{d})_0+
(\mathbf{a}\mathbf{c})_\nu(\mathbf{b}\mathbf{d})_0
\right.\\
\label{eq:Stilde}
&&\qquad\qquad\qquad\left.+(\mathbf{b}\mathbf{c})_\nu(\mathbf{a}\mathbf{d})_0
\right]\;,
\end{eqnarray}
$\mu$ and $\nu$ stand for different combinations of "$+$" and "$-$",
and simplified subscripts $1,2,3$ and $4$ correspond to the sets of subscripts $(\omega,k)$, $(\omega_1,i)$, $(\omega_2,p)$, and $(\omega_{\mu\nu},s)$, respectively.

\section{Third harmonic generation}
\label{sec:thg}

The derived set of equations for modal amplitudes, Eqs.~(\ref{eq:modal_coupled}), takes into full account material and geometrical dispersion of linear and nonlinear coefficients. Numerical propagation within this model is a challenging task due to the need to compute double integrals in the r.h.s. at each step of an iteration procedure. Below we focus on the problem of third harmonic generation (THG), whereby a relatively narrow band-width pump in a particular mode at frequency $\omega_0$ generates signal in a (generally different) mode at the triple frequency $3\omega_0$. While the efficiency of any inter-modal nonlinear coupling strongly depends on the phase matching, cf. $e^{i\Delta\beta z}$ factors in the r.h.s. of Eq.~(\ref{eq:modal_coupled}), excitation of any other modes can be safely disregarded. Thus we can omit the mode indexes and treat the structure as being effectively single-mode. Furthermore, we split the amplitude function $A(\omega)$ into the pump and third harmonic (TH) parts:
\begin{equation}
A(\omega)=A_1(\omega-\omega_0)+A_3(\omega-3\omega_0)\;,
\end{equation}
assuming that $A_1(\omega)$ and $A_3(\omega)$ each are localized functions with a band-width $\Delta\omega\ll\omega_0$. Under this assumption, the integral in the r.h.s. of Eq.~(\ref{eq:modal_coupled}) is non-zero only in certain narrow frequency intervals. Within each of these intervals, we can neglect frequency dependence of nonlinear coefficients, replacing them with constants. Specifically, the following set of nonlinear coefficients is important for our case:
\begin{eqnarray}
\gamma_1&=&\gamma^{(++)}(\omega_0,\omega_0,\omega_0,\omega_0)\;,\\
\gamma_3&=&\gamma^{(++)}(3\omega_0,3\omega_0,3\omega_0,3\omega_0)\;,\\
\gamma_{13}&=&\gamma^{(++)}(\omega_0,3\omega_0,3\omega_0,\omega_0)\;,\\
\nonumber
\widetilde{\gamma}_{13}&=&\gamma^{(+-)}(\omega_0,\omega_0,\omega_0,3\omega_0)\\
&=&\gamma^{(-+)}(3\omega_0,\omega_0,\omega_0,\omega_0)\;.
\label{eq:gamma13_tilde}
\end{eqnarray}

Using Taylor expansions of $\beta(\omega)$ in a vicinity of the pump and TH frequencies:
\begin{eqnarray}
\beta(\delta=\omega-j\omega_0)\approx \beta_{0j}+\beta_{1j}\delta+\frac{1}{2}\beta_{2j}\delta^2\;, \;\; j=1,3\;,\qquad
\end{eqnarray}
introducing pulse envelope functions:
\begin{eqnarray}
\Psi_{j}(z,t)=\frac{1}{\sqrt{2\pi}}\int_{-\infty}^{+\infty}A_{\delta,j}
e^{i(\beta-\beta_{0j}-\alpha_j-\delta/v_g)z-i\delta t}d\delta\;,\qquad
\label{eq:pulse_envelopes}
\end{eqnarray}
where $A_{\delta,j}=A_j(\delta=\omega-j\omega_0)$, $v_g$ is a reference group velocity, and taking inverse Fourier transform of Eqs.~(\ref{eq:modal_coupled}),  the following set of coupled equations is obtained \cite{Gorbach2015,agrawal}:
\begin{eqnarray}
\nonumber
\partial_z\Psi_1&=&\left(\beta_{11}-\frac{1}{v_g}\right)\partial_t\Psi_1
-i\frac{\beta_{21}}{2}\partial^2_t\Psi_1 -\kappa_1\Psi_1\\
\nonumber
&&+ i\left(\gamma_1 |\Psi_1|^2 + 2\gamma_{13}|\Psi_3|^2\right)\Psi_1\\
\label{eq:THG_coupled_psi1}
&&+i3\widetilde{\gamma}_{13}\left(\Psi_1^*\right)^2\Psi_3e^{i\Delta\beta z}\;,\\
\nonumber
\partial_z\Psi_3&=&\left(\beta_{13}-\frac{1}{v_g}\right)\partial_t\Psi_3
-i\frac{\beta_{23}}{2}\partial^2_t\Psi_3 -\kappa_3\Psi_3 \\
\nonumber
&&+ i\left(\gamma_3 |\Psi_3|^2+2\gamma_{13}|\Psi_1|^2\right)\Psi_3\\
&&
+i\widetilde{\gamma}_{13}\Psi_1^3e^{-i\Delta\beta z}\;,
\label{eq:THG_coupled_psi3}
\end{eqnarray}
where $\Delta\beta=\beta_{03}+\alpha_3-3(\beta_{01}+\alpha_1)$.

As follows from the definition of pulse envelope functions in Eq.~(\ref{eq:pulse_envelopes}), and normalization of modal amplitudes $A_\delta$ in Eq.~(\ref{eq:energy_W}), the energy carried by pump and TH pulses is given by $W_j=\int_{-\infty}^{+\infty}|\Psi_j|^2 dt$, such that $|\Psi_j|^2$ gives power.

In the limiting case of a continuous wave (CW) pump and a weak TH signal, $|\Psi_3|\ll|\Psi_1|$, neglecting pump depletion, nonlinear shift of the pump propagation constant ($\sim \gamma_1 |\Psi_1|^2$) and cross-phase modulation ($\sim \gamma_{13}$) terms, Eq.~(\ref{eq:THG_coupled_psi3}) reduces to the first-order linear ODE \cite{agrawal}:
\begin{equation}
\partial_z\Psi_3=-\kappa_3\Psi_3+i\widetilde{\gamma}_{13}\Psi_1^3e^{-i\Delta\beta z}\;.
\label{eq:THG_reduced}
\end{equation}

For the input pump power $P_1$, $\Psi_1(z)=\sqrt{P_1}e^{-\kappa_1 z}$, and zero input in the TH component, $\Psi_3(0)=0$, the solution of Eq.~(\ref{eq:THG_reduced}) can be written in terms of the {\em THG efficiency} $\eta$:
\begin{eqnarray}
\label{eq:eta_vs_z_CW}
\eta=\frac{|\Psi_3|^2}{P_1}=P_1^2 |\widetilde{\gamma}_{13}|^2 
\frac{\left|\sin\left(\Delta\widetilde{\beta} z/2\right)\right|^2}{\left(|\Delta\widetilde{\beta}|/2\right)^2}e^{-(\kappa_3+3\kappa_1) z}\;,\qquad\\
\Delta\widetilde{\beta}=\Delta\beta+i(\kappa_3-3\kappa_1)
=\Delta\beta+i\Delta\kappa\;.\qquad
\end{eqnarray}

In contrast to the well-known zero attenuation limit \cite{agrawal}, the attenuations of pump ($\kappa_1$) and third harmonic ($\kappa_3$) enforce exponential decay of the generated TH signal at large distances. The optimal distance $z_0$, at which the maximum intensity of generated signal is observed, satisfies the following condition:
\begin{equation}
\textrm{Re}\left\{\frac{\Delta\widetilde{\beta}}{2}\tan^{-1}
\left(\frac{\Delta\widetilde{\beta}z_0}{2}\right)\right\}=\frac{\kappa_3+3\kappa_1}{2}\;.
\end{equation}
In particular, for the case of phase matching $\Delta\beta =0$, this gives:
\begin{equation}
z_0=\frac{1}{\Delta\kappa}\log\frac{\kappa_3}{3\kappa_1}\;,
\label{eq:z0}
\end{equation}
and the corresponding maximum efficiency per unit pump power is:
\begin{equation}
\frac{\eta_{max}}{P_1^2}=|\widetilde{\gamma}_{13}|^2 \frac{1}{3\kappa_1\kappa_3}
\left(\frac{3\kappa_1}{\kappa_3}\right)^{\frac{\kappa_3+3\kappa_1}{\Delta\kappa}}\;.
\label{eq:eta_max}
\end{equation}

\section{An example: graphene-coated dielectric micro-fibre}
\label{sec:thg_fibre}

\begin{figure*}
\includegraphics[width=0.8\textwidth]{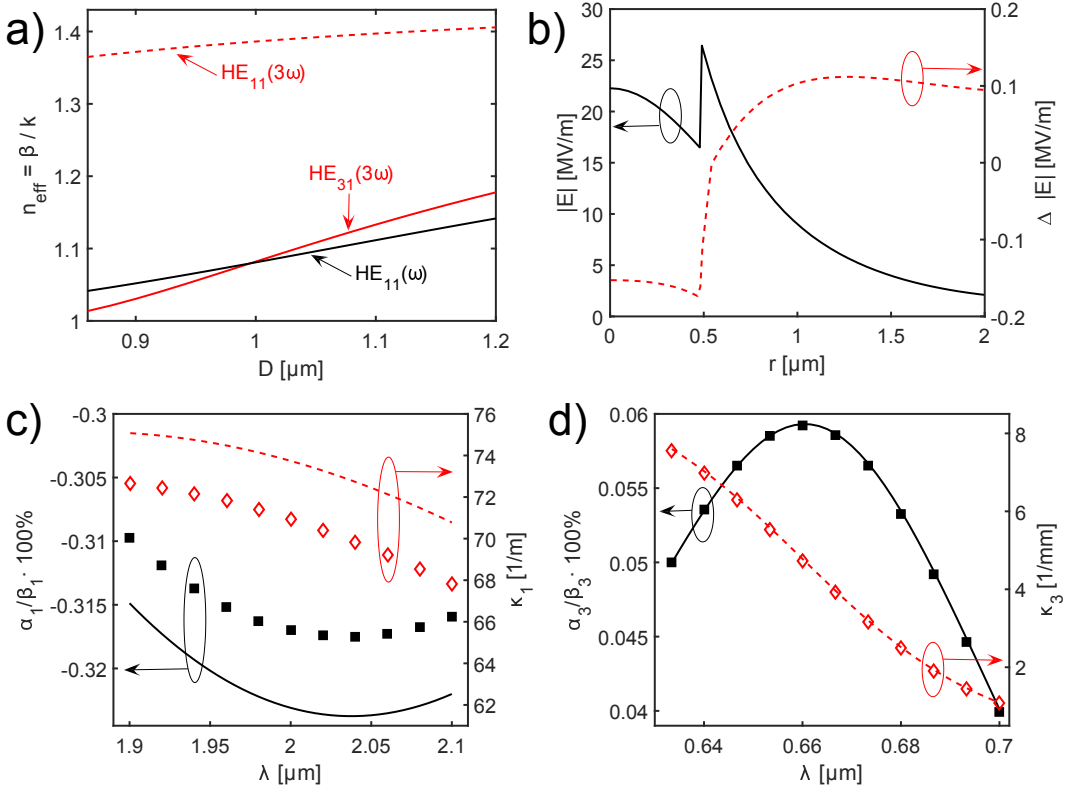}
\caption{(Color online) Graphene-coated silica micro-fibre: a) effective indexes of pure fibre modes (without the coating) at the fundamental frequency $\omega$ (black) and third harmonic $3\omega$ (red/grey) as functions of the fibre diameter $D$, $\lambda=2\pi c/\omega=2\mu m$; b) Profile of the pure fibre $HE_{11}$ mode at the fundamental frequency, $P_1=1W$, $\lambda=2\mu m$. Red/grey dashed curve indicates corrections to the shape of the mode when the fibre is fully coated with graphene; c) and d) corrections to the propagation constants and attenuation constants in fully coated fibre for the fundamental $HE_{11}$ (solid black) and third harmonic $HE_{31}$ (dashed red/grey) modes as functions of the fundamental harmonic wavelength, as computed from Eq.~(\ref{eq:kappa}). Solid suares and open diamonds indicate corresponding values computed with the help of FEM Maxwell solver Comsol. In b)-d) Fermi level of graphene is set to $E_F=0.93$eV.}
\label{fig:fibre_charact}
\end{figure*}

A graphene coated dielectric fibre represents one simple example of graphene integrated photonic structures \cite{Gao2014,Wu2015, Gorbach2013ol}. Profiles and propagation constants of all guided modes in a step-index dielectric fibre can be obtained semi-analytically \cite{agrawal}. Here we consider a fibre with silica glass core and air cladding. For small enough fibre diameters one can achieve phase matching between the fundamental guided mode ($HE_{11}$) at frequency $\omega$ and a higher order mode at frequency $3\omega$ \cite{Grubsky2005,Coillet}. For the pump wavelength $\lambda_0=2\pi c/\omega_0=2\mu$m, the phase matching with the $HE_{31}$ higher mode is achieved when the fibre diameter is $D\approx 0.98\mu m$, see Fig.~\ref{fig:fibre_charact}(a). While the diameter of such micro-fibres is comparable to the wavelength of the fundamental mode ($D\approx \lambda_0/2$), considerable field overlaps with a graphene coating of the fibre core can be achieved, cf. Fig.~\ref{fig:fibre_charact}(b). Therefore one can benefit from a graphene-induced boost of the effective nonlinearity of the structure \cite{Gorbach2013ol}. 

Recent theoretical analysis suggests that the third-order nonlinear graphene conductivity $\sigma_3(\omega_0,\omega_0,\omega_0,3\omega_0)$ responsible for THG process, see Eq.~(\ref{eq:gamma13_tilde}) , is resonantly enhanced at $\hbar\omega_0=2 E_F/3$, where $E_F\equiv |\mu|$ is the Fermi energy of graphene \cite{Mikhailov2015}. For $\lambda_0=2\mu m$ this gives $E_F\approx 0.93$eV. In a vicinity of the resonance, the nonlinear conductivity can be approximated as \cite{Mikhailov2015}:
\begin{eqnarray}
\nonumber
&&\sigma_3(\omega_0,\omega_0,\omega_0,3\omega_0)=\\
\label{eq:sig33}
&&\qquad-\sigma_0^{(3)}\frac{3}{32}\frac{E_F\hbar\tau^{-1}}
{[\hbar(\omega_0+i\tau^{-1}/3)-2E_F/3]^2}\;,\\
&&\sigma_0^{(3)}=\frac{e^4\hbar v_F^2}{4\pi E_F^4}\;,
\label{eq:sig03}
\end{eqnarray}
where $v_F=10^6$m/s is Fermi velocity, and $\tau$ is the phenomenological relaxation time.
Linear graphene conductivity is given by \cite{Falkovsky2008}:
\begin{eqnarray}
\nonumber
&&\sigma_1=\frac{i2e^2k_BT}{\pi\hbar^2(\omega+i\tau^{-1})}\ln\left[2\cosh\left(\frac{E_F}{2k_BT}\right)\right]\\
\label{eq:sigma1}
&&
+\frac{e^2}{4\hbar}\left[G\left(\frac{\omega}{2}\right)+i\frac{2\omega}{\pi}\int_0^{+\infty}\frac{G(\omega^\prime/2)-G(\omega/2)}{\omega^2-(\omega^\prime)^2}d\omega^\prime\right]\;,\qquad\\
\nonumber
&&G(\omega)=\frac{\sinh\left[\hbar\omega/(k_BT)\right]}
{\cosh\left[E_F/(k_BT)\right]+\cosh\left[\hbar\omega/(k_BT)\right]}\;.
\end{eqnarray}

In our simulations we set $\tau=200$fs \cite{Yan2013b} and room temperature $T=300$K. For this relaxation time, the nonlinear conductivity in Eq.~(\ref{eq:sig33}) reaches the peak value of $|\sigma_3|_{max}\approx 2.7\cdot 10^{-21} Sm^2/V^2$ for $E_F=0.93$eV and $\lambda_0=2\mu$m.

It is easy to see that all linear and nonlinear graphene coefficients in Eqs.~(\ref{eq:kappa}), (\ref{eq:gam_gr_pm})-(\ref{eq:gam_gr_pp}) are proportional to the fraction $f$ of the fibre core surface area coated with graphene: $0<f=L/(\pi D)<1$, where $L$ is the length of graphene contour in the cross-section of the structure \cite{Gorbach2013ol} (cf. also subsection 2 of the Appendix). Remarkably, this implies that the maximal THG efficiency does not depend on $f$, but the corresponding propagation length $z_0$ scales linearly with $f$, see Eqs.~(\ref{eq:eta_max}) and (\ref{eq:z0}). For simplicity, below we set $f=1$, i.e. we assume that the entire circumference of the fibre core edge is homogeneously coated with a single layer graphene. 

To analyze an influence of the graphene coating on the profiles of guided modes, we simulated the structure in the commercial finite element method Maxwell solver Comsol Multiphysics, where graphene was modeled as surface current. It was found that graphene induces only minor corrections to the shapes of fundamental and higher order modes, see Fig.~\ref{fig:fibre_charact}(b). This allows us to set $\hat{\sigma}_l=0$ in the perturbation expansion analysis, cf. Eq.~(\ref{eq:Jl}), and hence use modes of the uncoated micro-fibre when calculating coefficients in Eqs.~(\ref{eq:kappa}), (\ref{eq:gam_gr_pm})-(\ref{eq:gam_gr_pp}).

In Fig.~\ref{fig:fibre_charact}(c) and (d) we plot the graphene-induced corrections to the propagation constants and attenuation constants of the fundamental and TH modes, respectively. The results of our perturbation theory given by Eq.~(\ref{eq:kappa}) are in good agreement with the corresponding values computed directly from Comsol simulations. The discrepancies are more pronounced (but still remain as low as few percent) in the fundamental harmonic: at larger wavelengths fibre modes are less localized, and the graphene coating induces stronger perturbations to the shape of the modes.

\begin{figure}
\includegraphics[width=0.45\textwidth]{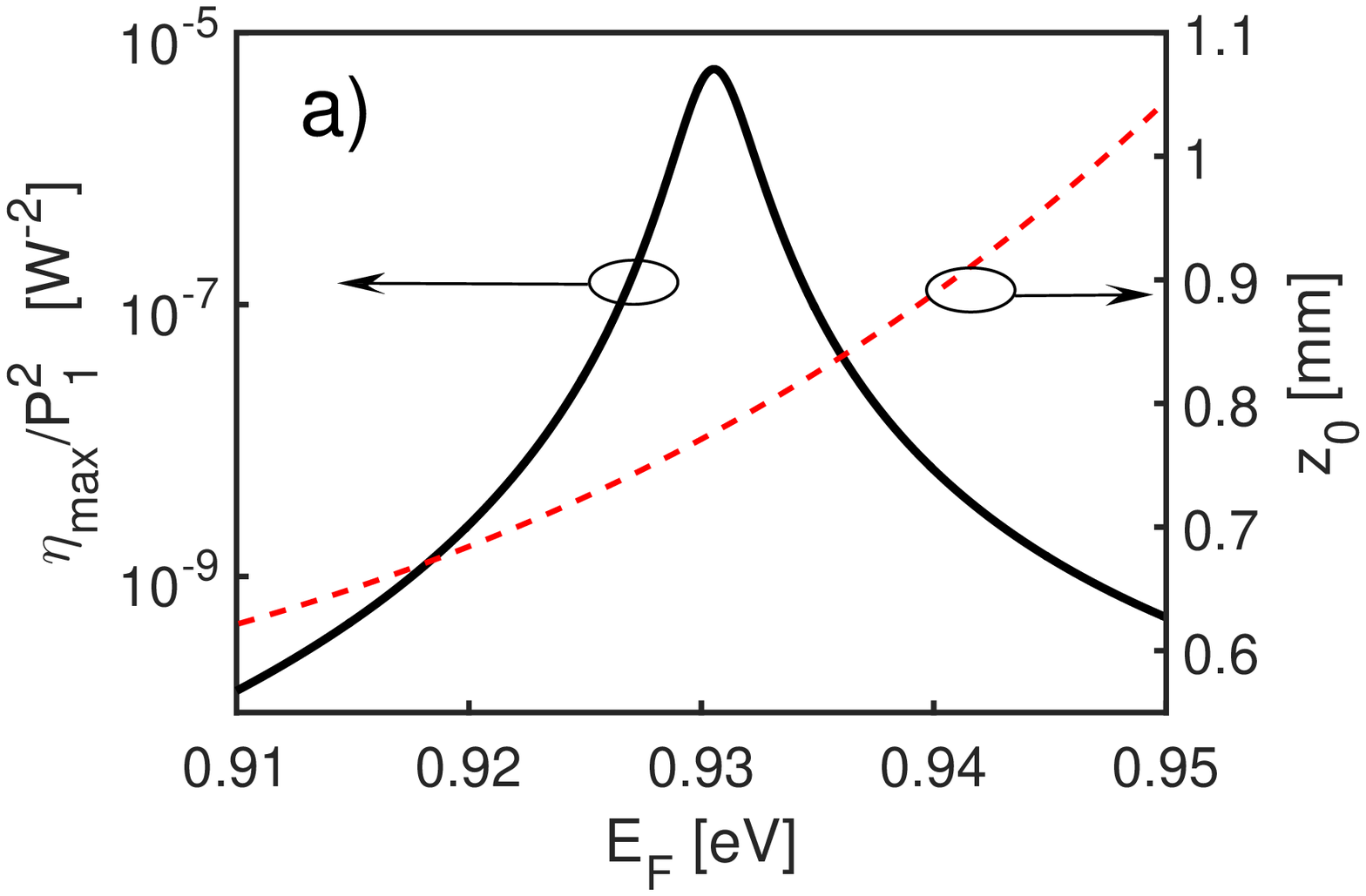}
\includegraphics[width=0.45\textwidth]{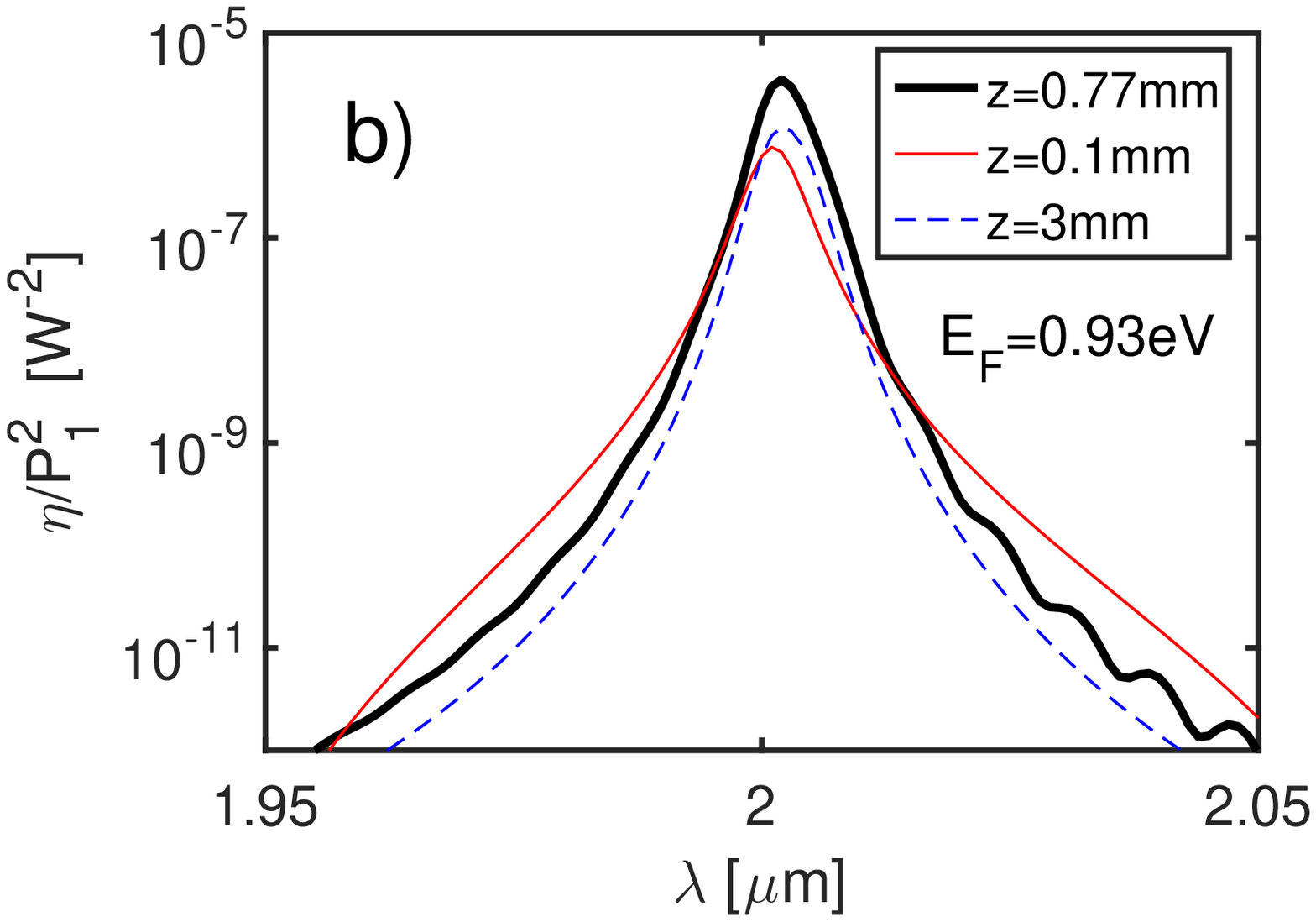}
\caption{(Color online) THG efficiency in a graphene coated micro-fibre of the diameter $D=0.98\mu$m: a) maximal efficiency and the optimal length $z_0$ for the case of pure phase matching $\Delta\beta=0$ (corresponding pump wavelength $\lambda\approx 2\mu$m); b) efficiency for different propagation distances as function of pump wavelength, $E_F=0.93$eV.}
\label{fig:eta_for_fibre}
\end{figure}

At resonance $\hbar\omega_0=2 E_F/3$, graphene is found to give by far the strongest contribution
to the overall nonlinear coefficient $\tilde{\gamma}_{13}$: $|\tilde{\gamma}_{13}^{(g)}|/|\tilde{\gamma}_{13}^{(d)}|\sim 2500$ (for silica glass fibre core we take $\chi_3=1.73\cdot 10^{-22} m^2/V^2$ \cite{agrawal}). The resulting maximal THG efficiency per unit pump power can be as high as $\eta_{max}/P_1^2=5\cdot 10^{-6} W^{-2}$, and the corresponding optimal propagation distance is below $1$mm (for the case of a fully coated fibre, $f=1$), see Fig.~\ref{fig:eta_for_fibre}(a). In Fig.~~\ref{fig:eta_for_fibre}(b) the efficiency is plotted as function of the pump wavelength when the graphene Fermi level is fixed at $E_F=0.93$eV, and for different propagation distances. The bandwidth of efficient frequency conversion is determined by the interplay between the resonance width of nonlinear conductivity and the dispersion of fibre modes. Remarkably, deviations from the optimal propagation distance $z_0\approx 0.77$mm within a considerably wide range from $z=0.1$mm to $3$mm reduce the THG efficiency by no more than one order of magnitude. 

\begin{figure}
\includegraphics[width=0.45\textwidth]{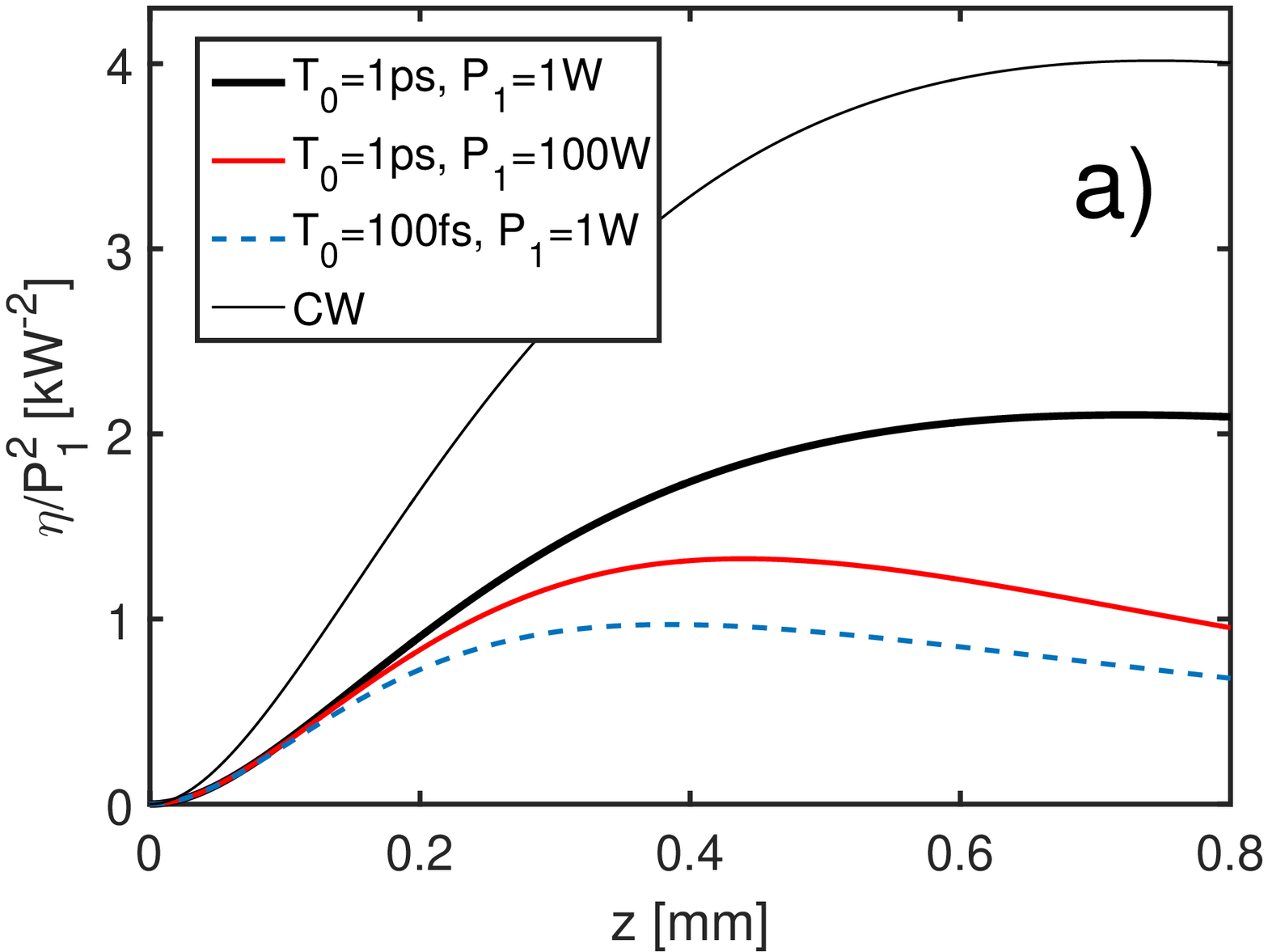}
\includegraphics[width=0.45\textwidth]{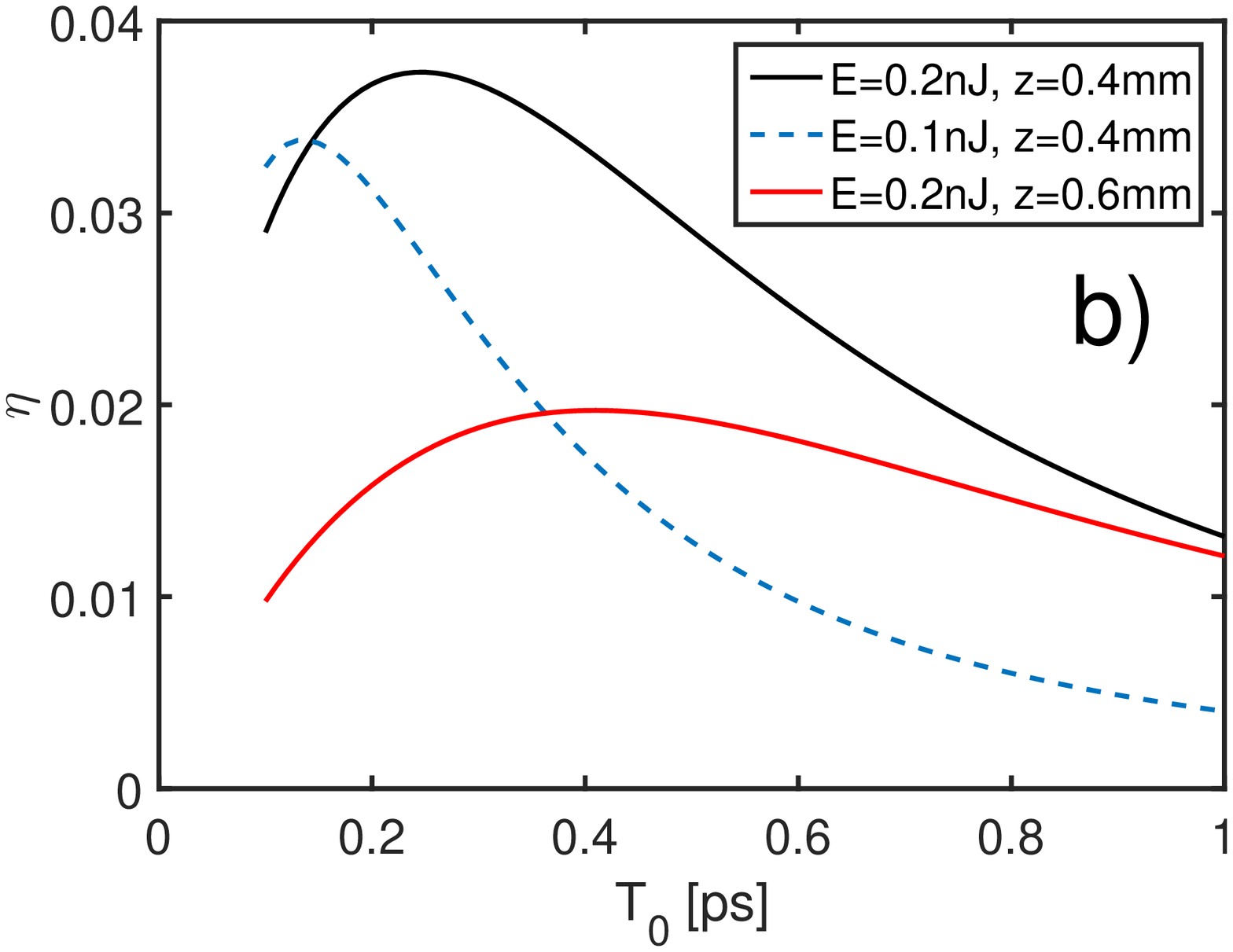}
\caption{(Color online) THG efficiency with a pulse excitation: a) normalized efficiency for different input peak powers and pulse duration, thin solid line indicates efficiency for the CW case, cf. Eq. (\ref{eq:eta_vs_z_CW}); b) efficiency as function of the input pulse duration for different levels of input pump energy $W_1(z=0)=E$ and a fixed propagation distance. Fibre and input parameters: $D=0.98\mu$m, $E_F=0.93$eV, $\lambda_0=2\mu$m.}
\label{fig:eta_for_fibre_pulses}
\end{figure}

To analyze THG process with a pulse pump excitation, we numerically solve Eqs.~(\ref{eq:THG_coupled_psi1}), (\ref{eq:THG_coupled_psi3}) with the initial condition: $\Psi_1(z=0)=\sqrt{P_1}\textrm{sech}(t/T_0)$, $\Psi_3(z=0)=0$. It is convenient to determine the THG efficiency as the ratio of pulse energies in this case: $\eta(z)=W_3(z)/W_1(0)$. 

We set $E_F=0.93\mu$m, $D=0.98\mu$m and $\lambda_0=2\mu$m, such that $\Delta\beta=0$, and $\tilde{\gamma}_{13}$ is resonantly enhanced at the pump central wavelength. To calculate graphene contribution to $\gamma_{11}$ nonlinear coefficient, we adopt the low-frequency approximation for the nonlinear conductivity: $\sigma_3(\omega_0,\omega_0,\omega_0,\omega_0)= -i (3/8) (E_F/\hbar\omega)^3 \sigma_0^{(3)}\approx 1.4\cdot 10^{-23} Sm^2/V^2$ \cite{Mikhailov2008, Mikhailov2015}. While the amplitude of the generated TH signal is relatively low, $|\Psi_3|^2\ll |\Psi_1|^2$, all terms with the two remaining nonlinear coefficients $\gamma_{13}$ and $\gamma_{33}$ are confirmed to have no noticeable impact on the THG process and can be safely disregarded.

The dispersion coefficients for the fundamental and third harmonics are calculated to be $\beta_{21}=12.8 ps^2/m$ and $\beta_{23}=5.2 ps^2/m$ respectively, and the group velocity mismatch is $\beta_{13}-\beta_{11}\approx 745$ps/m. For pulses of duration $T_0>1$ps the characteristic walk-off length $L_W=T_0/|\beta_{13}-\beta_{11}|$ and the dispersion lengths $L_D=T_0^2/|\beta_{2j}|$ ($j=1,3$) are all larger than the predicted optimal propagation distance $z_0$, cf. dashed line  in Fig.~\ref{fig:eta_for_fibre}(a). In this regime, and for low peak powers $P_1$, the THG efficiency follows the analytical result in Eq.~(\ref{eq:eta_vs_z_CW}) obtained for CW pump (up to a scaling factor due to different definitions of $\eta$ in these two cases), cf. thick and thin black curves in Fig.~\ref{fig:eta_for_fibre_pulses}(a). Reducing the pulse duration to $T_0=100$fs, the peak THG efficiency drops and is achieved at a shorter distance $z<z_0$ due to the walk-off between the pump pulse and generated TH signal, see dashed curve in Fig.~\ref{fig:eta_for_fibre_pulses}(a). 

For large peak powers $P_1$, the self-phase modulation of the pump induces a considerable effective phase mismatch ($\Delta\beta \sim \gamma_{11}P_1$). This effect counter-balances the growth of THG efficiency with the square of peak power $\eta\sim P_1^2$ predicted in Eqs.~(\ref{eq:eta_vs_z_CW}) and (\ref{eq:eta_max}). For $T_0=1$ps the increase of peak power from $P_1=1$W to $P_1=100$W reduces the maximal normalized efficiency $\eta_{max}/P_1^2$ by a factor of $\sim 1.5$, compare black and red/gray curves in Fig.~\ref{fig:eta_for_fibre_pulses}(a). In addition, the optimal distance, at which the maximum of $\eta$ is observed, reduces when increasing the peak power. Therefore, when comparing THG efficiency at a fixed propagation distance, the deviation from the simple parabolic law $\eta\sim P_1^2$ can become even more pronounced.

Due to the combination of the above walk-off and self-phase modulation effects, for a given energy of the input pump $E=W_1(z=0)$, there is an optimal pulse duration, and the corresponding peak power $P_1=E/(2T_0)$, which give the maximal efficiency at a fixed distance, see Fig.~\ref{fig:eta_for_fibre_pulses}(b). Reducing the pulse duration, and hence increasing its peak power, the efficiency grows initially, but it drops again when the optimal THG distance for the high peak power and short pulse becomes much shorter than the fixed length of the structure. Remarkably, the predicted efficiency of few percent is by many orders of magnitude larger than typical efficiency obtained in un-coated silica fibres ($\eta\sim 10^{-7}$) \cite{Coillet} and highly nonlinear nano-plasmonic waveguides ($\eta\sim 10^{-4}$) \cite{Sederberg2015}.

When the band-width of a short input pulse becomes comparable to the width of resonance of the nonlinear conductivity, see Eq.~(\ref{eq:sig33}), the dispersion of nonlinearity starts to play an important role. In this regime, the reduced model in Eqs.~(\ref{eq:THG_coupled_psi1}), (\ref{eq:THG_coupled_psi3}) is no longer applicable, and the appropriate analysis of the THG process can be done within the coupled modes model in Eq.~(\ref{eq:modal_coupled}). The corresponding studies are beyond the scope of the present work.

\section{Summary}

Using perturbation expansion of Maxwell equations with nonlinear polarization and surface current terms, we derived the coupled modes model in Eq.~(\ref{eq:modal_coupled}) which can be applied for analysis of generic third-order nonlinear frequency mixing processes in graphene integrated waveguides. This model takes into full account dispersions of linear and nonlinear conductivity of graphene, as well as susceptibilities of bulk materials. For a particular case of third harmonic generation from a narrow band-width pump, and assuming that the phase matching condition is satisfied for a specific pair of the fundamental and third harmonic guided modes, the above model is reduced to a conventional set of coupled nonlinear Schr{\"{o}}dinger type Eqs.~(\ref{eq:THG_coupled_psi1}), (\ref{eq:THG_coupled_psi3}).

We applied the derived models for the analysis of third harmonic generation in a graphene coated dielectric micro-fibre. Considering graphene induced corrections to the guided modes' propagation and attenuation constants, we demonstrated that the predicted values from our perturbation analysis are in good agreement with those obtained numerically with the help of the commercial Maxwell solver package. We also predicted the extraordinary high third harmonic generation efficiency of up to few percent from a $0.1$nJ sub-picosecond pump in a sub-millimeter long graphene coated fibre, when operating near the resonance $\hbar\omega=(2/3)E_F$ of the graphene nonlinear conductivity.

The described in our work perturbation theory can also be applied for analysis of nonlinear optical phenomena in structures containing other emerging 2D materials \cite{Castellanos-Gomez2016}.

\begin{acknowledgements}
Financial support from the Rank Prize Fund is gratefully acknowledged.
\end{acknowledgements}

\appendix

\section{Details of the perturbation expansion procedure}

\subsection{Planar graphene interface}

Consider a waveguide structure with integrated planar graphene ribbon located at $x=0$, $-L/2<y<L/2$. 
The linear operator in Eq.~(\ref{eq:operator_L}) can be written as:
\begin{equation}
\hat{L}=\left[
\begin{array}{ccc}
q^2-\partial^2_y & \partial^2_{xy} & i\beta\partial_x \\
\partial^2_{xy} & q^2-\partial^2_x & i\beta\partial_y \\
i\beta\partial_x & i\beta\partial_y & -\epsilon k^2-\partial^2_x-\partial^2_y
\end{array}
\right]\;,
\end{equation}
where $q^2=\beta^2-\epsilon k^2$, $k=\omega/c$. The boundary conditions for the mode $\mathbf{e}_\omega$ are:
\begin{eqnarray}
\Delta[e_{z}]=0\;,\qquad \Delta[e_{y}]=0\;,\\
\Delta[\partial_y e_{x}-\partial_x e_{y}]=-\frac{i\omega}{\epsilon_0 c^2}\left(\hat{\sigma}_l\mathbf{e}\right)_y\;,
\label{eq:BC_o12_1}
\\
\Delta[i\beta e_{x}-\partial_x e_{z}]=-\frac{i\omega}{\epsilon_0 c^2}\left(\hat{\sigma}_l\mathbf{e}\right)_z\;.
\label{eq:BC_o12_2}
\end{eqnarray}

In the order $O(s^3)$ of the perturbation expansion, Eq.~(\ref{eq:Os32}) is obtained with the operator $\hat{M}$ defined as:
\begin{eqnarray}
\hat{M}=\left[
\begin{array}{ccc}
2i\beta & 0 & -\partial_x \\
0 & 2i\beta & -\partial_y \\
-\partial_x & -\partial_y & 0
\end{array}
\right]\;,
\label{eq:M_cart}
\end{eqnarray}
and the boundary conditions:
\begin{widetext}
\begin{eqnarray}
\Delta[B_{z}]=0\;,\qquad \Delta[B_{y}]=0\;,\qquad\\
%\nonumber
\sum_j\left\{\Delta[\partial_y B_{jx}-\partial_x B_{jy}]+\frac{i\omega}{\epsilon_0c^2}\left(\hat{\sigma}_lB_j\right)_y\right\}e^{i\beta_j z}=
%\qquad\\
-\frac{i\omega}{\epsilon_0c^2 }J_{py}\;,
\label{eq:BC_o32_1}
\qquad\\
%\nonumber
\sum_j\left\{\Delta\left[\frac{\partial_zA_\omega}{\sqrt{N_\omega}} e_{x}+i\beta B_{x}-\partial_x B_{z}\right]
%\right.\qquad\\
%\nonumber
%\left.
+\frac{i\omega}{\epsilon_0c^2}\left(\hat{\sigma}_lB_j\right)_z\right\}e^{i\beta_j z}=
%\qquad\\
-\frac{i\omega}{\epsilon_0c^2 }J_{pz}\;.\qquad
\label{eq:BC_o32_2}
\end{eqnarray}

Computing projections of different terms in Eq.~(\ref{eq:Os32}) with the mode $\mathbf{e}_k$, we split integrals in $x$ as $\int_{-\infty}^{+\infty} dx= \int_{-\infty}^0 dx + \int_0^{+\infty} dx$ and take integrals by parts to obtain:
\begin{eqnarray}
\nonumber
&&\left<\mathbf{e}_k|\hat{L}|\mathbf{B}_j\right>=\left<\mathbf{B}_j|\hat{L}|\mathbf{e}_k\right>^*+\\
%\\
%\nonumber
%&&\qquad
&&\int_{-L/2}^{L/2}\left\{B_{jz}\Delta\left[i\beta_ke_{kx}^*+\partial_x e_{kz}^*\right]+
%\right.\\
%\nonumber
%&&\qquad\qquad
B_{jy}\Delta\left[\partial_xe_{ky}^*-\partial_y e_{kx}^*\right]+
e_{kz}^*\Delta\left[i\beta_jB_{jx}-\partial_xB_{jz}\right]
+e_{ky}^*\Delta\left[\partial_y B_{jx}-\partial_x B_{jy}\right]
\right\}dy\;,
\label{eq:L_not_adjoint}
\qquad\\
&&\left<\mathbf{e}_k|\hat{M}|\mathbf{e}_k\right>=i(\beta_j+\beta_k)\iint_{-\infty}^{+\infty}
\left(\mathbf{e}_k^*\cdot\mathbf{e}_j\right)
dxdy
-\int_{-L/2}^{L/2}\Delta\left[e_{kx}^* e_{jz}\right]dy\;.
\end{eqnarray}
Applying boundary conditions from Eqs.~(\ref{eq:BC_o12_1}), (\ref{eq:BC_o12_2}), (\ref{eq:BC_o32_1}), (\ref{eq:BC_o32_2}) we therefore derive:
\begin{eqnarray}
\nonumber
&&\sum_j\left\{
\left<\mathbf{e}_k|\hat{L}|\mathbf{B}_j\right>-\frac{1}{\sqrt{N_j}}\partial_z A_j
\left<\mathbf{e}_k|\hat{M}|\mathbf{e}_k\right>
\right\}e^{i\beta_j z}=\\
\nonumber
&&\qquad
\sum_j-\frac{i\partial_z A_j}{\sqrt{N_j}}\left\{
(\beta_j+\beta_k)
\iint_{-\infty}^{+\infty}
\left(\mathbf{e}_k^*\cdot\mathbf{e}_j\right)
dxdy+i\int_{-L/2}^{L/2}\Delta\left[e_{kx}^*e_{jz}-e_{kz}^*e_{jx}\right]dy
\right\}e^{i\beta_j z}\\
\label{eq:a11}
&&
\qquad\qquad\qquad\qquad
-\frac{i\omega}{\epsilon_0c^2}\int_{-L/2}^{L/2}
\left(\mathbf{e}_k^*\cdot\mathbf{J}_p\right)dy\;.\;
\end{eqnarray}

Re-writing the normalization condition in Eq.~(\ref{eq:Iw}) in terms of electric field only, applying integration by parts, and using the relationship $\textrm{div}(\mathbf{e})=0$, it is possible to show that:
\begin{eqnarray}
(\beta_j+\beta_k)
\iint_{-\infty}^{+\infty}
\left(\mathbf{e}_k^*\cdot\mathbf{e}_j\right)
dxdy+i\int_{-L/2}^{L/2}\Delta\left[e_{kx}^*e_{jz}-e_{kz}^*e_{jx}\right]dy=\delta_{jk}\frac{4\omega N_j}{\epsilon_0 c^2}\;,
\end{eqnarray}
and thus we obtain Eq.~(\ref{eq:after_appendix}).

\end{widetext}

\subsection{Structures with radial symmetry}

Consider a radially-symmetric waveguide with a graphene ribbon located along an arc of radius $R$, spanning the angle $0<\phi<\Phi$ ($0<\Phi<2\pi$). Adopting cylindrical coordinates, the linear operator in Eq.~(\ref{eq:operator_L}) is:
\begin{equation}
\hat{L}=\left[
\begin{array}{ccc}
q^2-\frac{\partial^2_\phi}{r^2} & \frac{\partial_\phi}{r^2}\partial_r r & i\beta\partial_r\\
\partial_\phi\partial_r\frac{1}{r} & q^2-\partial_r\frac{1}{r}\partial_r r & \frac{i\beta}{r}\partial_\phi\\
\frac{i\beta}{r}\partial_r r & \frac{i\beta\partial_\phi}{r} & -p^2-\frac{\partial^2_\phi}{r^2}-\partial_r \frac{1}{r}\partial_r r
\end{array}
\right],
\end{equation}
where $q^2=\beta^2-\epsilon k^2$, $p^2=1/r^2+\epsilon k^2$, and the boundary conditions are:
\begin{eqnarray}
\Delta[e_{z}]=0\;,\qquad \Delta[e_{\phi}]=0\;,\\
\frac{1}{R}\Delta[\partial_\phi e_{r}-r\partial_r e_{\phi}]=0\;,\\
\Delta[i\beta e_{r}-\partial_r e_{z}]=0\;,
\end{eqnarray}

In the order $O(s^3)$ of the perturbation expansion, the operator $\hat{M}$ in Eq.~(\ref{eq:Os32}) is:
\begin{eqnarray}
\hat{M}=\left[
\begin{array}{ccc}
2i\beta & 0 & -\partial_r \\
0 & 2i\beta & -\frac{\partial_\phi}{r} \\
-\frac{1}{r}\partial_r r & -\frac{\partial_\phi}{r} & 0
\end{array}
\right]\;,
\label{eq:M_cyl}
\end{eqnarray}
and the boundary conditions are:
\begin{widetext}
\begin{eqnarray}
\Delta[B_{z}]=0\;,\qquad \Delta[B_{\phi}]=0\;,\qquad\\
%\nonumber
\sum_j\left\{\frac{1}{R}\Delta[\partial_\phi B_{jr}-r\partial_r B_{j\phi}]+\frac{i\omega}{\epsilon_0c^2}\left(\hat{\sigma}_lB_j\right)_\phi\right\}e^{i\beta_j z}=
%\qquad\\
-\frac{i\omega}{\epsilon_0c^2 }J_{p\phi}\;,
\label{eq:BC_o32_1_cyl}
\qquad\\
%\nonumber
\sum_j\left\{\Delta\left[\frac{\partial_zA_\omega}{\sqrt{N_\omega}} e_{r}+i\beta B_{r}-\partial_r B_{z}\right]
%\right.\qquad\\
%\nonumber
%\left.
+\frac{i\omega}{\epsilon_0c^2}\left(\hat{\sigma}_lB_j\right)_z\right\}e^{i\beta_j z}=
%\qquad\\
-\frac{i\omega}{\epsilon_0c^2 }J_{pz}\;.\qquad
\label{eq:BC_o32_2_cyl}
\end{eqnarray}
\end{widetext}

Following the same procedure as described in the previous section, and splitting integration in radial coordinate as: $\int_0^{+\infty}dr=\int_0^Rdr+\int_R^{+\infty}dr$, we derive the equation which is similar to Eq.~(\ref{eq:a11}) but with all line integrals replaced as: $\int_{-L/2}^{L/2}(\dots)dy \to \int_0^\Phi (\dots) Rd\phi$, and field components $e_{x}$ replaced by $e_{r}$ in the argument of $\Delta$ function.

Generalizing the above results onto the case of an arbitrary shaped graphene contour integrated into a waveguide cross-section, we split the contour into infinitesimally small planar and arc sections, and hence obtain the generic result in Eq.~(\ref{eq:modal}).

\bibliography{joint_bib}

\end{document}